\documentclass[prd,amsfonts,twocolumn,nofootinbib,showpacs]{revtex4}
\RequirePackage{graphicx}
\usepackage{enumerate}

\begin{document}
\title{Ambiguity in running spectral index with an extra light field during inflation}

\author{Kazunori Kohri}
\affiliation{Cosmophysics group, Theory Center, IPNS, KEK,  
and The Graduate University for Advanced Study (Sokendai), Tsukuba 305-0801, Japan}
\author{Tomohiro Matsuda}
\affiliation{Laboratory of Physics, Saitama Institute of Technology, Fukaya, Saitama 369-0293, Japan}

\begin{abstract}
At the beginning of inflation there could be extra dynamical scalar fields
that will soon disappear (become static) before the end of inflation.
In the light of multi-field inflation, those extra degrees of freedom
may alter the time-dependence of the original spectrum of the curvature
 perturbation.
It is possible to remove such fields introducing extra number of e-foldings
 prior to $N_e\sim 60$, however such extra e-foldings may make the
 trans-Planckian problem worse due to the Lyth bound.
We show that such extra scalar fields can change the running of the
 spectral index to give correction of $\pm 0.01$ without adding
 significant contribution to the spectral index.
The corrections to the spectral index (and the amplitude) could be
 important in considering global behavior of the corrected spectrum,
 although they can be neglected in the estimation of the spectrum and
 its spectral index at the pivot scale. 
The ambiguity in the running of the spectral index, which could be
due to such fields, can be used to nullify tension between BICEP2 and
 Planck experiments.
\end{abstract}

\maketitle

\section{Introduction}
The running of the spectral index has sometimes been used to discriminate
inflationary models.
Recent discovery of B-mode polarization by the BICEP2
collaboration~\cite{Ade:2014xna} would be an important signal of
primordial inflation, since it would indicate the existence of quantum
gravitational radiation, which is a generic prediction of inflationary
cosmological models~\cite{Lyth:1998xn, Olive:1989nu}.\footnote{See also
Ref.~\cite{Mortonson:2014bja}, which shows (in contrast to 
Ref.~\cite{Ade:2014xna}) that joint BICEP2
$+$ Planck analysis could favor solutions without gravity waves.} 
Although the tensor perturbation is very powerful in discriminating
inflationary models, the BICEP2 result is in some tension with previous
experiments such as WMAP~\cite{Hinshaw:2012aka} and
Planck~\cite{Ade:2013uln}, which claimed upper limits on $r$ (tensor to
scalar ratio).
To resolve the tension, BICEP2 collaboration
suggested~\cite{Ade:2014xna} that the scalar spectral index could run
fast.
However, the signature of the required running $\alpha_s=-0.022\pm 0.010$
($68\%$CL)~\cite{Ade:2014xna} is negative and its absolute value is very
large compared with previous expectation.\footnote{See also recent studies~\cite{recent-st}}
In this paper we show that the constraints from the running spectral
index could be nullified by introducing an extra free scalar field.
Multifield models of inflation may allow inflationary trajectory that is
sensitive to the initial condition.
Recently, it was pointed out that some multifield models make
predictions for the inflationary observables that do not depend strongly
on the specific initial condition~\cite{multi-stat, Easther:2013rva}.
On the other hand, observational data could be used to constrain the
initial field configuration if the model has sensitivity to the
initial condition. 
Recent numerical calculation~\cite{Easther:2013rva} clearly shows that
such sensitivity can generate significant running in 
double quadratic inflation.\footnote{In this paper we investigate a
possibility of shifting $\alpha_s$ without changing significantly the other
cosmological predictions.
Although Fig.3 of Ref.\cite{Easther:2013rva} is
illuminating, it might be showing that the
model could not give $\alpha_s\sim -0.01$ without shifting $n_s$.}
Our result might indicate that the running spectral index could be a free
parameter that may not be taken too much seriously in this setup.

\section{Running spectral index with a free scalar field}
\subsection{A toy model: Extra component $\dot{\rho}_\chi/\rho_\chi=$const.}
We consider a model in which the curvature
perturbation is sourced by the inflaton $\phi$.
The additional free scalar field is $\chi$, which is dynamical at the
beginning of inflation and may contribute to the adiabatic curvature
perturbation at that moment.

In this section we introduce extra field $\chi$, which will soon
disappear because of $\dot{\rho}_\chi =-6H\rho_\chi<0$.
Note that this toy model has $\ddot{\rho}_\chi>0$, whose contributeion to 
the running is usually $\Delta \alpha_s>0$.
Therefore, although the model is simple and very useful for the intuitive
argument, the model is not suitable for generating negative running.

The curvature perturbation at the horizon exit is calculated as
\begin{eqnarray}
\label{zeta-ini}
\zeta&=&-H\frac{\delta
 \rho}{\dot{\rho}}\nonumber\\ 
&=& -H\frac{\delta \rho_\phi+\delta
 \rho_\chi}{\dot{\rho}_\phi+\dot{\rho}_\chi}\nonumber\\  
&\equiv& r_\phi \zeta_\phi +(1-r_\phi)\zeta_\chi,
\end{eqnarray}
where $\rho_i$ and $P_i$ are the density and the pressure of the component
labeled by $i$.
For the inflaton $\phi$, they are defined as
\begin{eqnarray}
\rho_\phi&\equiv&\frac{1}{2}\left(\dot{\phi}\right)^2
+ V(\phi)\nonumber\\
P_\phi&\equiv&\frac{1}{2}\left(\dot{\phi}\right)^2- V(\phi).
\end{eqnarray}
Here $\rho\equiv \rho_\phi+\rho_\chi$ is defined for the total energy density.
We used component perturbation $\zeta_i$ and the ratio $r_\phi$, which are
defined by
\begin{eqnarray}
\zeta_i&\equiv& -H\frac{\delta \rho_i}{\dot{\rho}_i},\nonumber\\
r_\phi&\equiv& \frac{\dot{\rho}_\phi}{\dot{\rho}_\phi+\dot{\rho}_\chi}.
\end{eqnarray}
If $\chi$ has a flat potential $V(\chi)=0$ and initial velocity
$v_{\chi0}\ne 0$, it will have a decaying velocity
$\dot{\chi}(t)=v_{0}e^{-3Ht}$ and could have a negligible perturbation $\delta \rho_\chi$.
The energy density of $\chi$ obeys $\dot{\rho}_\chi=-3H(1+w)\rho_\chi$
$(w=1)$.
For our calculation we define a ratio between densities: $f\equiv
\rho_\chi/\rho_\phi$, for which inflationary expansion requires $f\ll 1$.
One might argue that $\zeta_\phi$ is not conserved after horizon exit, since the
equation of motion does depend on $\rho_\chi$ via the Hubble parameter 
$H^2= \frac{\rho}{3M_p^2}=\frac{\rho_\phi+\rho_\chi}{3M_p^2}$.
The equation of motion is
\begin{eqnarray}
\ddot{\phi}+3H\dot{\phi}+V_{,\phi}=0,
\end{eqnarray}
where the subscript with comma means derivatives with respect to the
field.
Conservation of $\zeta_\phi$ is expected when $P_\phi$ is practically a
unique function of $\rho_\phi$~\cite{New-cons}.
In our scenario, $P_\phi$ might not be an exact unique function of
$\rho_\phi$, however for $f\ll 1$, one can expect that both $P_\phi$ and
$\rho_\phi$ are practically unique functions of $\phi(t)$.
Barring small correction from $f\ll 1$, the curvature perturbation after 
inflation will be $\zeta\simeq \zeta_\phi$~\cite{Infcurv}, which has the
spectrum 
\begin{eqnarray}
\label{curv-pert-spe}
{\cal P}_{\zeta}&=&
 \left(\frac{H}{\dot{\phi}}\right)^2\left(\frac{H}{2\pi}\right)^2
\nonumber\\
&=&\frac{1}{8\pi^2M_p^2}\frac{H^2}{\epsilon_\phi},
\end{eqnarray}
where the slow-roll parameter is defined by
\begin{eqnarray}
\epsilon_\phi &\equiv& \frac{M_p^2}{2}\left(\frac{V_{,\phi}}{\rho}\right)^2.
\end{eqnarray}
Here, $\zeta_\phi$ is defined at the horizon exit.
For later calculation we also define
\begin{eqnarray}
\eta_\phi&\equiv& \frac{V_{,\phi\phi}}{3H^2}.
\end{eqnarray}
We are assuming that the energy density $\rho_\chi$ of the additional
field is a negligible fraction of the total energy density during inflation.
Although $\rho_\chi$ is a small fraction of the total energy density,
$|\dot{\rho}_\chi|\sim |\dot{\rho}_\phi|$ is possible when $\phi$ is
slow-rolling.
(If we make an assumption that our mechanism does not change the original
$n_s-1$, we need to consider $|\dot{\rho}_\chi|\lesssim |\dot{\rho}_\phi|$.
In that case, contribution to $\alpha_s$ will be maximum when 
$|\dot{\rho}_\chi|\sim  |\dot{\rho}_\phi|$.
We are not fine-tuning the quantities to obtain
$|\dot{\rho}_\chi|\sim  |\dot{\rho}_\phi|$, since we have little
difficulty with obtaining the required
$\alpha_s$.)
Moreover, one can easily imagine that $|\ddot{\rho}_\chi|\gg
|\ddot{\rho}_\phi|$ could be possible even if $f\ll 1$.
Our idea is based on that simple speculation.

Since we are introducing a free scalar field $\chi$, $\epsilon_H$ is
splitted into two parts:
\begin{eqnarray}
\epsilon_H&\equiv& -\frac{\dot{H}}{H^2}\nonumber\\
&=&-\frac{\dot{\rho}_\phi}{6H^3 M_p^2}
-\frac{\dot{\rho}_\chi}{6H^3 M_p^2}\nonumber\\
\nonumber\\
&\equiv&\epsilon_{\phi}+\epsilon_{H\chi}.
\end{eqnarray}
Introducing $R\equiv \rho_\chi/3H^2M_p^2$, we find
\begin{eqnarray}
\epsilon_{H\chi} &=& \frac{3(1+w)}{2}R.
\end{eqnarray}
Here $\epsilon_{H\chi}$ should be discriminated from $\epsilon_\chi$,
which will be used in later calculation.
$\epsilon_{H\chi}$ is identical to $\epsilon_{\chi}$ only when $\ddot{\chi}$ is
negligible.
Then, using $d\ln k =Hdt$, we find for the slow-rolling inflaton:
\begin{eqnarray}
\frac{1}{d\ln k}\epsilon_\phi
&=&4\epsilon_\phi\epsilon_H-2\eta_\phi\epsilon_\phi.
\end{eqnarray}
In the first term, $\epsilon_H$ appears instead of $\epsilon_\phi$, since
our definition of $\epsilon_\phi$ uses $\rho$ instead of $V$.
For our calculation we also evaluate
\begin{eqnarray}
\frac{1}{d\ln k}\epsilon_{H\chi}&=&
\frac{3(1+w)}{2H}\dot{R}
\nonumber\\
&=&\frac{3(1+w)}{2H}\left[R\frac{\dot{\rho}_\chi}{\rho_\chi}-2R\frac{\dot{H}}{H}\right]
\nonumber\\ 
&=&-3(1+w)\epsilon_{H\chi}+2\epsilon_{H\chi}\epsilon_H.
\end{eqnarray}
The spectral index of the curvature perturbation is 
\begin{eqnarray}
\label{eq-index}
n_s-1&\equiv&\frac{d\ln {\cal P}_\zeta}{d\ln k}\nonumber\\
&=&-6\epsilon_H+2\eta_\phi.
\end{eqnarray}
Since $\epsilon_H=\epsilon_\phi+\epsilon_{H\chi}$, the shift caused by
the additional field $\chi$ is 
\begin{eqnarray}
\Delta (n_s-1)&=&-6\epsilon_{H\chi}.
\end{eqnarray}
Since we are considering a conservative scenario in which additional field $\chi$ does not
change the original index, we need $6\epsilon_{H\chi}\ll 0.04$.
We also have 
\begin{eqnarray}
\frac{1}{d\ln
 k}\eta_\phi&=&\frac{1}{Hdt}\left[\frac{V_{,\phi\phi}}{3H^2}\right]\nonumber\\
&=&\frac{1}{H}\left(\frac{V_{,\phi\phi\phi}\dot{\phi}}{3H^2}\right)+2
\left(\frac{V_{,\phi\phi}}{3H^2}\right)\left(-\frac{\dot{H}}{H^2}\right)\nonumber\\
&=&-\left(\frac{V_{,\phi\phi\phi}V_{,\phi}}{9H^4}\right)+2
\left(\frac{V_{,\phi\phi}}{3H^2}\right)\left(-\frac{\dot{H}}{H^2}\right)\nonumber\\
&=&-\xi_\phi +2\eta_\phi\epsilon_H,
\end{eqnarray}
where we defined
\begin{eqnarray}
\xi_{\phi}&\equiv& \frac{V_{,\phi} V_{,\phi\phi\phi}}{9H^4}.
\end{eqnarray}
Then, the running of the spectral index is
\begin{eqnarray}
\label{alpha-eq-kin}
\alpha_s&\equiv&\frac{d n_s}{d\ln k}
\nonumber\\
&=& -6\left[4\epsilon_\phi\epsilon_H-2\eta_\phi\epsilon_\phi\right]\nonumber\\
&&+2\left[2\eta_\phi\epsilon_H -\xi_\phi\right]\nonumber\\
&&-6\left[-3(1+w)\epsilon_{H\chi}+2\epsilon_{H\chi}\epsilon_H\right]
\nonumber\\
&\simeq& \left[-24\epsilon_\phi\epsilon_H+16\eta_\phi\epsilon_\phi -2\xi_\phi\right]
\nonumber\\
&&+4\eta_\phi\epsilon_{H\chi}+18(1+w)\epsilon_{H\chi}-12\epsilon_{H\chi}\epsilon_H,
\end{eqnarray}
where we can see correction $\Delta \alpha_s\sim 18(1+w)\epsilon_{H\chi}\lesssim
0.12(1+w)$, but the sign of this term is positive. 
We will see that the correction can be negative when $\chi$ is moving on
a hilltop potential.

For our scenario, we are considering an inflationary model in which
final $\zeta$ is determined by $\phi$, but an extra
component may change scale dependence of the spectrum.
Multi-field models of inflation have been studied by many
authors~\cite{Polarski:1992dq, Bassett:2005xm, Vernizzi:2006ve}.
From those studies one can see that an additional field, which could be
dynamical at the horizon exit but will soon stop or
slow down, may not change final $\zeta$ at least at the first order.
The usual calculation uses adiabatic and entropy perturbations instead of
component perturbations, and considers conversion between them.
Although in this paper we will consider
calculation that is usually considered for the curvaton
mechanism~\cite{Infcurv}, one will reach the same result using the calculation
considered in Ref.\cite{Polarski:1992dq, Bassett:2005xm,
Vernizzi:2006ve}.
A similar idea has been studied when there is no extra field but there
is a deviation from the 
slow-roll velocity~\cite{Damour:1997cb, Liddle:1998pz,
Taruya:1998cz, Yokoyama:1998rw, Seto:1999jc, Takamizu:2010je}.
In those studies it has been found that the curvature perturbation could
not be affected by such additional degrees of freedom if they are
disappearing during inflation.
Also in Ref.~\cite{matsuda-elliptic, Matsuda:2008fk, Matsuda:2012kk},
it has been pointed out that such field might shift the spectral index and
its running. 
Viewing those studies, we think that the idea of changing the scale
dependence of the curvature perturbation using an additional dynamical
field could have been known partly, although we could not find explicit
calculation that can be used for solving the tension between BICEP2 and
other experiments.

\subsection{slow-roll $\chi$}
In the above calculation we considered an additional component that
obeys $\dot{\rho}_\chi=-3H(1+w)\rho_\chi$.
Let us see what happens if $\chi$ is a moderately slow-rolling field
(but it will soon reach its minimum or will soon be negligible).
Using conventional slow-roll parameters for both $\phi$ and $\chi$, 
we find the spectral index
\begin{eqnarray}
\label{eq-index2}
n_s-1&=&-6\epsilon_H+2\eta_\phi,
\end{eqnarray}
where $\epsilon_H\equiv \epsilon_\phi+\epsilon_\chi$.
Here we define $\epsilon_\chi\equiv
\frac{M_p^2}{2}\frac{V_{,\chi}}{\rho}$, which is equivalent to
$\epsilon_{H\chi}$ when $\ddot{\chi}$ is negligible.
The running of the spectral index is
\begin{eqnarray}
\label{alpha-eq}
\alpha_s
&=& -6\left[4\epsilon_\phi\epsilon_H-2\eta_\phi\epsilon_\phi\right]\nonumber\\
&&+2\left[2\eta_\phi\epsilon_H -\xi_\phi\right]\nonumber\\
&&-6\left[4\epsilon_\chi\epsilon_H-2\eta_\chi\epsilon_\chi\right]
\nonumber\\
&\simeq& \left[-24\epsilon_\phi\epsilon_H+16\eta_\phi\epsilon_\phi -2\xi_\phi\right]
\nonumber\\
&&+4\eta_\phi\epsilon_\chi-24\epsilon_\chi\epsilon_H
+12\eta_\chi\epsilon_\chi,
\end{eqnarray}
where the major correction appears in $|12\eta_\chi\epsilon_\chi|.$
If we assume $|\Delta (n_s-1)|=6\epsilon_\chi<|n_s-1|$, we will 
have $|\Delta \alpha_s|<|2\eta_\chi
(n_s-1)|\sim 0.08|\eta_\chi|$.
Since $\eta_\chi$ does not appear in the spectral index, there is no
bound for $|\eta_\chi|$.
Therefore, one can take $|\eta_\chi|$ as large as $0.125$, which gives 
correction as large as $|\Delta \alpha_s|\sim 0.01$. 
Since we are considering an extra dynamical field that will soon
disappear,
$\eta_\chi\sim 0.125$ is a conceivable choice for the model.
Since $\eta_\chi<0$ is possible for a hilltop potential, one will be able to find
a correction $\Delta \alpha_s \sim -0.01$. 

\subsection{fast-roll $\chi$}
Alternatively, one can consider a hilltop potential $V(\chi)=
 -\frac{1}{2}m_\chi^2 \chi^2$ for a fast-rolling field $\chi$ and will
 find~\cite{Dimopoulos:2003ce}  
\begin{eqnarray}
\chi(t)&=& \chi(t_0)e^{Kt},
\end{eqnarray}
where 
\begin{eqnarray}
K&\equiv&
 \frac{3}{2}H\left[1-\sqrt{1-\frac{4}{9}\left(\frac{m_\chi^2}{H^2}\right)} 
\right].
\end{eqnarray}
Conventional fast-rolling condition is ``$\epsilon_\chi\ll
1$ but $|\eta_\chi|\sim {\cal O}(1)$''.
Then the slow-roll parameter can be replaced as
\begin{eqnarray}
\epsilon_{H\chi}&=&-\frac{\dot{\rho}_\chi}{6H^3M_p^2}\nonumber\\
&=&\frac{(-K^3+K m_\chi^2)\chi^2 }{6H^3M_p^2}\nonumber\\
&=&\left(-\frac{K^2}{m_\chi^2}+1\right)\frac{K}{H}R,
\end{eqnarray}
where contribution from the kinetic term has not been neglected since
$\ddot{\chi}$ could not be negligible for the fast-rolling field.
Then we find 
\begin{eqnarray}
\frac{1}{d\ln k}\epsilon_{H\chi}&=&
-\frac{\ddot{\rho}_\chi}{6H^4M_p^2}
+3\epsilon_{H\chi}\epsilon_H
\nonumber\\
&=&\frac{2K}{H}\epsilon_{H\chi}
+3\epsilon_{H\chi}\epsilon_H.
\end{eqnarray}
The spectral index is 
\begin{eqnarray}
n_s-1&=&-6\epsilon_\phi+2\eta_\phi
-6\epsilon_{H\chi},
\end{eqnarray}
which gives $6\epsilon_{H\chi}<0.04$ for our scenario.
The running of the spectral index is
\begin{eqnarray}
\alpha_s&=&-6[4\epsilon_\phi\epsilon_H-2\eta_\phi\epsilon_\phi]
+2[2\epsilon_H\eta_\phi-\xi_\phi]\nonumber\\
&&-6\left[\frac{2K}{H}\epsilon_{H\chi}
+3\epsilon_{H\chi}\epsilon_H\right]\nonumber\\
&=& \left[-24\epsilon_\phi\epsilon_H+16\eta_\phi\epsilon_\phi -2\xi_\phi\right]
+4\epsilon_{H\chi}\eta_\phi\nonumber\\
&&-6\left[\frac{2K}{H}\epsilon_{H\chi}
+3\epsilon_{H\chi}\epsilon_H\right].
\end{eqnarray}
The correction appears in  $-12\epsilon_{H\chi}K/H$, where $\Delta
\alpha_s$ is negative for the hilltop potential.
If one takes $K/H\sim 1$ (fast-roll), one will find 
$\Delta \alpha_s\sim -0.01$ for $\epsilon_{H\chi}\sim 8\times 10^{-4}$.

\subsection{The Curvaton}
For the conventional curvaton mechanism, one will
find~\cite{Kobayashi:2012ba} 
\begin{eqnarray}
\alpha_s&\simeq&2\frac{\ddot{H}}{H^3}-4\epsilon_H^2
+4\epsilon_H\eta_\sigma
-2\xi_\sigma,
\end{eqnarray}
where $\sigma$ is the curvaton.
If we introduce an extra light field $\chi$, significant correction may
appear from the first term, which gives for a fast-rolling $\chi$:
\begin{eqnarray}
2\frac{\ddot{H}}{H^3}&=& \frac{\ddot{\rho}_\chi}{3H^4M_p^2}+...\nonumber\\
&\sim&2\left(1-\frac{K^2}{m_\chi^2}\right)\frac{K^2}{H^2}R.
\end{eqnarray}
Therefore, one can easily find similar correction in various other models including
the curvaton.

\subsection{A model}
In this section we show a specific scenario in which $\alpha_s$ can be
 shifted by an additional field $\chi$.
At the beginning of inflation, the additional field has a hilltop potential 
\begin{eqnarray}
V(\chi)=-\frac{1}{2}m_\chi^2\chi^2,
\end{eqnarray}
and $\chi$ will soon reach the minimum at $\chi_0$.
Here we assume $\chi_0\sim M_p$.
The duration is bounded by $e^{K\Delta t}=\frac{\chi_\mathrm{end}}{\chi_\mathrm{ini}}<\frac{M_p}{H}\simeq
2.4\times 10^{4}$~\cite{Ade:2014xna}, 
where 
 $\chi_\mathrm{ini} > \delta \chi \simeq H/2\pi$ and
 $\chi_\mathrm{end}<\chi_0\sim M_p$ has been considered.
We find 
$\Delta N\lesssim 10\times\left(\frac{H}{K}\right)$.
The additional field $\chi$ must reach its minimum before the end of inflation.

For a moderately slow-rolling $\chi$, $\epsilon_{\chi}$ is determined by
$m_\chi$ and $H$.
Then $|\Delta (n_s-1)|<0.04$ gives for the quadratic potential:
\begin{eqnarray}
6\epsilon_\chi&=&6\times \frac{M_p^2}{2} \left(\frac{-m_\chi^2
					  \chi}{\rho}\right)^2_*
=6\times
\left(\frac{-m_\chi^2}{3H^2}\right)_*\left(\frac{-\frac{1}{2}m^2_{\chi}\chi^2}{\rho}\right)_*
 \nonumber\\
&=&6\eta_\chi R_*<0.04.
\end{eqnarray}
Here the quantities with star is defined
when the perturbation leaves horizon.
We consider $\eta_\chi\sim 0.6$ and $|R_*|\sim 0.01$ on the scale where
significant $\alpha_s$ could be observed.
The condition $|R_*|\sim 0.01$ is conceivable since it is required at the
very early stage of inflation.  
Then, we can estimate the correction  as
\begin{eqnarray}
|\Delta \alpha_s|&\simeq & |12\epsilon_\chi\eta_\chi|=|12\eta_\chi^2
 R|\sim 0.04.
\end{eqnarray}
In the same way, the running of $\alpha_s$ will be shifted by a similar
order, which is consistent with Planck
experiment.\footnote{See Fig.3 of Ref.~\cite{Ade:2013uln}.}

On the other hand, one might suspect that higher
runnings may ovecome the correction coming form $\alpha_s$.
The expansion will be~\cite{Kohri:2013mxa}
\begin{eqnarray}
{\cal P}_\zeta(k)&=&{\cal P}_\zeta(k_0)\exp\left[
(n_s-1)\ln\left(\frac{k}{k_0}\right)
+\frac{\alpha_s}{2}\ln^2\left(\frac{k}{k_0}\right)\right.\nonumber\\
&&\left.+\frac{\beta_s}{3!}\ln^3\left(\frac{k}{k_0}\right)
+...\right]\nonumber\\
&=&
{\cal P}_\zeta(k_0)
\left(\frac{k}{k_0}\right)^{
(n_s-1)
+\frac{\alpha_s}{2}\ln\left(\frac{k}{k_0}\right)
+\frac{\beta_s}{3!}\ln2\left(\frac{k}{k_0}\right)
+...}.
\end{eqnarray}
Note that there will be no problem in the expansion if one takes
      $\ln(k/k_0)=\ln(0.05/0.0027)\simeq 2.92$ and 
      $\beta_s < \alpha_s$, where $\beta_s$ is the running of $\alpha_s$.
Our model suggests that $\alpha_s$ and $\beta_s$ could be a similar
      order.
However, normally $\beta_s$ does not exceed $\alpha_s$, and the higher
      runnings does not
      spoil the scenario
 of nullifying the tension between BICEP2 and
      the Planck experiments.

We are anticipating that there could be some significant sign of new
physics in the higher runnings, however at this moment higher
      runnings are not so much constrained.
Such a large running (and  a running of running) will be checked by
the future B-mode polarization and the 21cm line
observations.~\cite{Kohri:2013mxa}.
One may use this to distinguish the source of the scale 
dependence if a signature might be observed in future
experiments.

\section{Conclusion and discussion}
\label{conclusion}
Normally in single-field inflation models, we have only one
scalar field that is responsible for both the inflationary expansion and
the production of the curvature perturbation.
In reality, however, there could be many other scalar fields {\it besides} the
inflaton field, which could be dynamical at least at the beginning of
the inflationary stage.
Although the non-inflaton components may disappear (or becomes static) 
before the end of inflation, those additional fields could change the
scale dependence of the spectrum.
If one wants to remove those extra fields before the onset of $N_e\sim
60$ inflation, one can assume extra e-foldings prior to $N_e\sim60$, so
that such fast-rolling (or modestly fast-rolling) fields are already 
negligible from the beginning.
In that case, remaining fields will not have significant
contribution to $\alpha_s$.  
On the other hand, in the light of BICEP2 and the Lyth bound, such extra e-foldings would
require more excursion of the inflaton field, which could be a problem
if one takes the trans-Planckian problem seriously.
We thus expect that such extra dynamical degrees of freedom can exist
naturally in modern inflationary scenarios. 
As long as we know, the idea has not been discussed yet to
solve the tension between BICEP2 and other experiments.

In this paper we considered a free scalar field and calculated its
contribution to $n_s$ and $\alpha_s$. 
We assumed conservatively that spectral index is not much altered by the
additional field.
Using a simple model, we showed that $O(0.01)$ ambiguity can be found in
the running spectral index if an extra field could be dynamical when the
perturbation leaves horizon.
In the same way, the running of $\alpha_s$ can be shifted by a similar
order, which could be used to distinguish the source of the scale 
dependence.
At the same time, the scale-dependence of the tensor mode could be
affected, which has already been discussed in
Ref.\cite{Gerbino:2014eqa, Gong:2014qga, Kobayashi:2013awa}.
In our scenario, we find $r\sim 16 \epsilon_\phi$, which does not
include $\epsilon_\chi$.
This means that $r$ is not affected by $\epsilon_\chi\ne 0$.
Similarly, the running of $r$ is determined by $\dot{\epsilon}_\phi$,
which does not introduce significant running at the first
order.
Significant running may appear from the next order of the running.\footnote{The second order (i.e, the running of of running of $r$)
can be large since it will include $\ddot{\epsilon}_\phi$.}
On the other hand, in our model $\frac{d n_t}{d\ln k}\propto \dot{\epsilon}_H$
can be large.
Since the scale-dependence of the tensor power spectrum is signicant
in our case, while such a scale dependence is usually not assumed,
there could be concern
that it might bias the estimation of parameters like $r$ and
$\alpha_s$  from the WMAP/Planck data.  Those quantities could be
negligible at the end, but the resolution of the tension between
BICEP2 and the WMAP/Planck could be affected.
In that sense we need further investigation to obtain more accurate
estimation of the parameters when the tensor mode is
running.\footnote{We thank the reviewer of the journal for these points.}

\section*{Acknowledgement}
K.K. is supported in part by Grant-in- Aid for Scientific
research from the Ministry of Education, Science, Sports, and
Culture (MEXT), Japan, Nos. 21111006, 23540327, 26105520 (K.K.). The
work of K.K. is also supported by the Center for the Promotion of
Integrated Science (CPIS) of Sokendai (1HB5804100).

\end{document}